\documentclass[iop]{emulateapj}  %  ApJ journal style

\def\ergcm2s{~erg cm$^{-2}$ s$^{-1}$ } % ergs per cm**2 second smay be
\def\ergs{~erg s$^{-1}$}% ergs per second 
\def\n4038{~NGC4038/39}% Use in text.  For title, use NGC 4038/4039
\def\x2{$\chi^{2}$}

\shorttitle{An accretion model for 4U 0142+61}
\shortauthors{Truemper  et al.}

\usepackage{graphicx}
\usepackage{epsfig}
\usepackage{apjfonts}
\usepackage{amsmath}


\begin{document}
\title{An accretion model for the anomalous X-ray pulsar 4U 0142+61}

\author{J. E. Truemper, K. Dennerl}
\affil{Max-Planck-Institut f\"{u}r extraterrestrische Physik, 
Postfach 1312, 85741 Garching, Germany}

\author{N. D. Kylafis}
\affil{
University of Crete, Physics Department \& Institute of Theoretical \& 
Computational Physics, 71003 Heraklion, Crete, Greece\\
Foundation for Research and Technology-Hellas, 71110 Heraklion, Crete,
Greece}

\author{\"{U}. Ertan}
\affil{
Faculty of Engineering and Natural Sciences, Sabanc\i\ University, 
34956, Orhanl\i, Tuzla, \.Istanbul, Turkey}

\author{A. Zezas}
\affil{
University of Crete, Physics Department \& Institute of Theoretical \& 
Computational Physics, 71003 Heraklion, Crete, Greece\\
Foundation for Research and Technology-Hellas, 71110 Heraklion, Crete,
Greece}

%%%\altaffiltext{1}{Visiting Astronomer, Cerro Tololo Inter-American Observatory.
%%%CTIO is operated by AURA, Inc.\ under contract to the National Science
%%%Foundation.}

%% editorial office after sub

\begin{abstract}

 We propose that the quiescent emission of AXPs/SGRs is powered by accretion from a fallback disk, requiring magnetic dipole fields in the range $10^{12}-10^{13}\,\rm{G}$, and that the luminous hard tails of their X-ray spectra are produced by bulk-motion Comptonization in the radiative shock near the bottom of the accretion column. This radiation escapes as a fan beam, which is partly absorbed by the polar cap photosphere, heating it up to relatively high temperatures. The scattered component and the thermal emission from the polar cap form a polar beam. We test our model on the well-studied AXP 4U 0142+61, whose energy-dependent pulse profiles show double peaks, which we ascribe to the fan and polar beams. The temperature of the photosphere (kT$\sim0.4$\,keV) is explained by the heating effect. 
The scattered part forms a hard component in the polar beam. We suggest that the observed high temperatures of the polar caps of AXPs/SGRs, compared with other young neutron stars, are due to the heating by the fan beam. Using beaming functions for the fan beam and the polar beam and taking gravitational bending into account, we fit the energy-dependent pulse profiles and obtain the inclination angle and the angle between the spin axis and the magnetic dipole axis, as well as the height of the radiative shock above the stellar surface. We do not explain the high luminosity bursts, which may be produced by the classical magnetar mechanism operating in super-strong multipole fields.
\end{abstract}


%________________________________________________________________

\section{Introduction}

Anomalous X-ray pulsars (AXPs) and Soft Gamma Ray Repeaters (SGRs) are young neutron stars with X-ray luminosities much larger than their spin-down power and long periods in the range
$2 - 12$ s. They are widely believed to be magnetars deriving their X-ray emission from the decay
of super-strong magnetic dipole fields ($B \gtrsim ~ 10^{15}$ G) (e.g.  Duncan \& Thompson 1992; Thompson
\& Duncan 1995).  At the outset, the magnetar model was developed to
explain the giant bursts and the large bursts of SGRs, which exceed
the Eddington limit of neutron-star luminosities by a very large
factor. If the observed spin-down of some of these sources is
interpreted as the consequence of magnetic dipole braking, the
resulting polar field strengths are of the order of  $\gtrsim 
10^{15}$, in qualitative agreement with what has been inferred from
the observed luminosities (Kouveliotou et al. 1998). Later on, it was
discovered that AXPs show short bursts as well, though less frequent
and less energetic than SGRs, leading to the general notion that both
types of sources are closely related or represent even a single
class. In the magnetar picture, the steady X-ray emission of these
sources, which have luminosities of typically a few times  $10^{35}$\,\ergs, is thought to be caused by a twist of the
magnetosphere leading to the amplification of the magnetic field and 
the acceleration of particles, which produce the X-ray emission.  The
twist is caused by rotational motions of a crust plate and has a
lifetime of $\sim 1$  year (Beloborodov \& Thompson 2007). For recent
reviews of the magnetar model see Woods \& Thompson (2006) and
Mereghetti (2008). 

An alternative energy source for the persistent and transient X-ray luminosities of AXPs and SGRs is accretion from fallback disks, first proposed by van Paradijs et al. (1995), and followed by Chatterjee et al. (2000) and Alpar (2001). This class of models was developed further in a series of papers (Ek\c{s}i \& Alpar 2003; Ertan \& Alpar 2003; Ertan \& Cheng 2004; 
Ertan et al. 2006; Ertan \& \c{C}al{\i}\c{s}kan 2006; Ertan et al. 2007; 
Ertan \& Erkut 2008, Ertan et al. 2009). The fallback-disk model gets support from the discovery of IR/optical radiation from two of the AXPs, 4U 0142+61 (Wang et al. 2006) and 1E 2259+586 (Kaplan et al. 2009), which has been successfully interpreted as disk emission. This optical emission is especially important, because the spectral fits
at short wavelengths put an upper limit on the inner disk radius and
thereby on the magnetic dipole field strength ( $\rm{B<10^{13}}$ G).

The fallback-disk model explains the spin-down of AXPs and SGRs by the
disk-magnetosphere interaction and requires ``normal'' neutron-star
dipole fields ($10^{12} - 10^{13}$ G). We note that this is in
line with the recent discovery of a magnetar (SGR 0418+5729) with a
low magnetic dipole field  $\rm{B < 7 \times 10^{12}}$  G (Rea et al. 2010; see also Alpar, Ertan \& {C}al{\i}\c{s}kan 2011). In addition, the model is successful in predicting the period clustering of AXPs/SGRs in the range of $2 - 12$ s. More recently, it was shown that the X-ray luminosity and the rotational properties, including the braking index $n=0.9\pm0.2$, of the so called ``high-B'' radio pulsar PSR J1734-3333 (Espinoza et al. 2011) can be explained by the evolution of a neutron star with a fallback disk and $B=10^{12}$ G (\c{C}al{\i}\c{s}kan et al., submitted). On the other hand, this model cannot explain the super-Eddington bursts, which are relatively rare. They are attributed to magnetar-type activities occurring in local {\textit{multipole}} fields (star-spots).

All references quoted above, regarding the fallback-disk model, are
dealing with the physics of the accretion disk and its interaction
with the magnetosphere of the neutron star. A first attempt to explain
the hard X-ray spectrum of the well-studied AXP 4U 0142+61 by
bulk-motion and thermal Comptonization has been presented by Truemper
et al. (2010). Fig. 1 shows the excellent fit of the COMPTB model to the pulse phase averaged data from Chandra MEG and HEG at low energies and Integral ISGRI above 15 keV.  In the present paper we study the energy-dependent
pulse profiles and the phase-dependent energy spectra, taking again 4U
0142+61 as an example.  

%Here goes Fig1

In \S\ 2 we summarize the observational properties of 4U 0142+61, in
\S\ 3 we present our model, in \S\ 4 we interpret the phase-dependent
energy spectra and infer the surface magnetic field strength, in \S\ 5
we discuss the heating of the polar cap, in \S\ 6 we compare 4U
0142+61 with low-luminosity X-ray pulsars in binary systems, and in
\S\ 7 we give our summary and conclusions.

\section{Observational properties of 4U 0142+61}

Since its discovery by Uhuru, 4U 0142+61 has been observed by many groups. The source has a
period of 8.69 s and spins down with a rate of $2 \times 10^{-12}$ s
s$^{-1}$  (Dib et al. 2007). It shows a spectrum
in the soft band (0.8 -- 10 keV) which can be fitted by a blackbody ($kT=0.42$ keV) and a steep
power law (photon-number spectral index $\Gamma_{s}=3.3$; e.g. Juett et al. 2002). At higher energies
($\rm{E>10}$ keV), a hard spectral tail ($\Gamma_{h}=0.93$) is observed, which extends beyond 150 keV and
contributes about one third of the total X-ray luminosity (den Hartog et al. 2008). That paper
presents the most detailed data on the spectra and light curves of the persistent soft and hard X-ray
emission of this source, based on observations with ASCA, XMM-Newton, RXTE, and
INTEGRAL. In our present analysis, we use the data provided by den Hartog et al. (2008), since
they give very detailed information on the energy-dependent pulse shapes (in particular their Fig.
9). According to this data, the luminosity of 4U 0142+61 in the energy band 0.8 -- 160 keV is
$\sim 4.6 \times 10^{35}$\,\ergs, assuming a distance of 3.6 kpc (Durant \& van Kerkwijk 2006). The
luminosity in the soft and the hard band is $\sim 3.2 \times 10^{35}$\,\ergs\ and $\sim 1.4 \times 10^{35}$\,\ergs\ respectively.

The broad-band spectrum of 4U 0142+61 has been measured by Suzaku (Enoto et al. 2011) in the
0.4 - 70 keV band, confirming the existence of a two-component spectrum. The photon index of
the hard component is $\Gamma_{h}=0.9$. The Suzaku observations have the advantage that data in the hard
($>10$ keV) and the soft (1 -- 10 keV) bands are taken simultaneously. The luminosities are
 $2.8 \times 10^{35}$\,\ergs\  in the soft band (1 -- 10 keV)
 and  $\sim 6.8 \times 10^{34}$\,\ergs\,  in the hard band (10-- 70 keV). We compared the luminosities derived from the Suzaku spectral fits with those from the
combined XMM-Newton and INTEGRAL spectral fits in the same bands and we found excellent
agreement (Table 1).

\begin{deluxetable}{lcccc}
\tablecolumns{5}
\tablewidth{0pc}
\tablecaption{Comparison between the data obtained with XMM-Newton/INTEGRAL and Suzaku}
\tablehead{
\colhead{Energy Range} & \colhead{XMM/Integral} & \colhead{Suzaku}\\
\colhead{(keV)} & \colhead{} & \colhead{} }
\startdata
 1 -- 10 keV   &  $2.8 \times 10^{35}$  &   $2.8 \times
 10^{35}$ \\
 10 --70 keV  &  $0.64 \times 10^{35}$  &  $0.64 \times
 10^{35}$ \\
\hline
 $\rm{kT_{blackbody}}$ (keV) &  $0.418\pm0.013^{\dag}$  & $0.428\pm0.004$ \\
Power-law index $\rm{\Gamma_{s}}$ & $3.3\pm0.4^{\dag}$ & $3.95\pm0.04$ \\
Power-law index $\rm{\Gamma_{h}}$ & $0.93$ & $0.11 - 1.54$ \\
\enddata
\tablecomments{$\dag$ We quote the Chandra data here (Juett et al. 2002), since den Hartog et al. (2008) did not perform
spectral fits to the low-energy data obtained with XMM-Newton.}
\end{deluxetable}  

The same holds for the spectral parameters, except in the case of the hard power-law index derived
from Suzaku, which strongly depends on the model for the soft component. In conclusion, this
comparison suggests that the quiescent spectrum of the source is quite stable. It also justifies our
choice to use the INTEGRAL data, which is of superior statistical
quality in the hard spectral range.

Using the observational data of den Hartog et al. (2008), we have derived the observed pulsed
profiles by adding the pulsed and non-pulsed components. The result is shown in Fig. 2. As noted
by den Hartog et al. (2008), morphology changes are ongoing throughout the whole energy range.
However, we stress that the principal structure of the pulse profiles is very simple: It mainly
consists of two pulses; one is located at phase $\sim0.1$, extending from the lowest to the highest
energies (0.8 - 160 keV). We call it the {\textit{main}} pulse. The {\textit{secondary pulse}} is located at phase $\sim0.6$.
Its relative strength decreases strongly from the lowest energy channel (0.8 - 2 keV) to medium
energies (8 -- 16.3 keV) and increases again in the (20 -- 50 keV)
band.

In Fig.~3 
we show the phase-dependent pulsed spectra derived from the data of den Hartog et al.
(2008). The spectrum of the main pulse is taken from the phase interval 0.85 -- 0.35 (shown as a dashed line),
while the spectrum of the secondary pulse is taken from the phase interval 0.35 -- 0.85 (shown as a dotted line).

%Here goes Fig2

\section{The general physical picture of our model}

The basic picture of our model is shown schematically in Fig. 4. It is based on the model which
has been developed by many authors, notably Davidson (1973), Basko \& Sunyaev (1976,
hereafter BS76) and Lyubarskii \& Sunyaev (1982, hereafter LS82). The accreting matter falls
freely on magnetic field lines until it is stopped in a radiative shock, where it is slowed down
from a supersonic to a subsonic flow in the sinking column beneath the shock. Most of the
accretion energy released in the radiative shock is transformed into hard X-ray photons by bulk-motion Comptonization (BMC, Blandford \& Payne 1981, Payne \& Blandford 1981, LS 82),
which escape sideways forming a fan beam. We note that in high luminosity sources, having a
large transverse optical depth in the accretion column  at the height of the shock, 
most of the BMC photons will be downscattered
before they can escape. As we discuss in section 3.2 below, due to the low luminosity of
AXPs and SGRs, the transverse optical depth is moderate ($\tau_{t}\sim5$ ). The height of the shock above
the stellar surface depends on the mass accretion rate and the cross section of the accretion
column. If the fan beam is formed very close to the neutron-star surface, then about half of it
escapes, while the other half - assisted by gravitational bending - hits the polar cap surrounding
the accretion column. The impinging flux heats the polar cap photosphere and some part of it is
scattered off by it. The scattered flux and the photospheric emission form a polar beam, which is
orthogonal to the fan beam. If the radiative shock forms at a height above the neutron-star
surface, then more than half of the fan beam escapes. The larger the height, the larger the
fraction of the fan beam that escapes.

\subsection{Outline of the model}

In the fallback-disk model, the magnetic dipole field strength of the neutron star is in the range
$10^{12} - 10^{13}$. Therefore, we can employ the radiation models developed for accreting highly
magnetized neutron stars in binary systems. Matter entering the magnetosphere of the neutron
 star at the Alfven radius falls down in the magnetic funnel provided by the dipole magnetic field
of the neutron star and the flow is converted from supersonic to subsonic at a radiative shock. In
the early works quoted above, a steady and homogeneous flow of matter was assumed, while
Morfill et al. (1984) were the first to discuss a time-dependent flow (accretion of clumps of
matter). Later on, Klein et al. (1996) studied non-steady accretion due to the formation of photon
bubbles. A fully consistent description of the complicated processes governing the flow and the
stopping of matter in the magnetic funnel is still lacking. Fortunately, however, the situation with
AXPs and SGRs can be expected to be much simpler, due to the significantly lower accretion rate
than that of most X-ray pulsars.

As discussed in \S\ 2, 4U 0142+61 shows basically a double-pulse structure, with different
spectra for the two components. We identify the {\textit{main pulse}}, which extends from low energies to
more than 100 keV, with the {\textit{fan beam}}, which is radiated sideways from the column (c.f. Fig. 4a,
dashed arrows), and whose high energy component ($> 10$ keV) is powered by BMC, taking place
in the radiative shock. In the process of BMC, seed photons from below the radiative shock gain
energy by back-and-forth collisions across the shock with the infalling, semi-relativistic electrons
and the slowly moving, post-shock, thermal plasma beneath the shock. The thermal energy of this
plasma is due to the electron recoil in the BMC process (LS82). It is well known that this first order
Fermi energetization produces a power-law spectrum.  

LS82 studied spectral formation due to bulk and thermal Comptonization in a radiation-dominated shock. In their model, the shock has a Rankine-Hugoniot form, and therefore the energy flux is strictly conserved. Of course, the shocks in actual X-ray pulsar accretion flows are radiative, but in practice, most of the radiation escapes through the column walls in the region downstream from the shock, and therefore the LS82 model may be an acceptable approximation for the structure of the shock itself. LS82 computed analytically the entire spectrum, not only the power law.
 On the other hand, the low energy (2 -- 10 keV)
steep power law of the main pulse may be attributed to thermal Comptonization (TC) taking place
in the sinking column, with seed photons produced by blackbody, bremsstrahlung, and cyclotron
processes. An external source of seed photons is the radiation from the photosphere surrounding
the accretion column.

%Figure 4 here

The {\textit{secondary pulse}} is attributed to photons emitted by the hot polar cap, forming a {\textit{polar}} beam
(dotted arrows in Fig. 4a. At low energies ($< 2$ keV), it consists of the photospheric black body-like
emission and at higher energies ($> 2$ keV) of fan beam photons, which are scattered when
they hit the photospheric plasma.

Accretion onto magnetic neutron stars has been studied extensively by Becker \& Wolff (2007;
hereafter BW07). With their model, BW07 studied spectral formation in X-ray pulsars and
highlighted the role of TC and BMC for the spectral formation in accretion columns. BW07
assumed a smooth flow in a cylindrical column, which is slowed down by a radiative shock. 
The shock is treated properly, not as a mathematical discontinuity, and it 
is the place where the BMC operates with seed photons produced by bremsstrahlung, cyclotron,
and blackbody processes inside the column. This model predicts a fan beam.  As we said above, we suggest that this process is
responsible for the high energy part of the spectrum ($E>10$ keV), which escapes as a fan beam.
Unfortunately, BW07 did not include an external source of seed photons. Nevertheless, their
model can be used to fit the spectrum of 4U 0142+61 at high energies, which is dominated by the
fan beam photons. We have done this by using the XSPEC version of the BW07 model provided
by Ferrigno et al. (2009) and find that it provides a good fit to the spectrum at energies > 10 keV
(Zezas et al., in preparation). However, fitting the soft part of the spectrum requires a thermally
Comptonized blackbody spectrum, which we attribute to reprocessing of the thermal radiation
from the polar cap.

Another relevant model for obtaining BMC and TC spectra has been developed by Farinelli et al.
(2008) for application to the spectra of low-mass X-ray binaries (LMXBs). It uses non-magnetic
(Thomson) cross sections, but for our purpose it has the advantage of including an external source
of seed photons. We used this model in our previous paper on 4U 0142+61 (Truemper et al. 2010)
and obtained a good fit to the broadband X-ray spectrum of the source (see Fig. 1). The fact that
it gives an excellent fit to both the soft and the hard components is significant, though the resulting
fit parameters would probably be somewhat different if magnetic cross sections were used. This
proviso applies to a lesser degree to the Ferrigno et al. (2009) model. For details of the application
of the Ferrigno et al. (2009) and Farinelli et al. (2008) models to 4U 0142+61 and other AXPs and
SGRs, we refer to Zezas et al. (in preparation).

\subsection{Physical parameters of the accretion column}

For simplicity, it is usually assumed that the accretion flow is homogeneous and the accretion
funnel is cylindrical. However, for detailed comparisons between model and observations of the
energy-depended pulse profiles, as we do below, this picture is too simple and qualitatively
inadequate. 

BS76 and LB82 have pointed out that, for a neutron star with an inclined magnetic field, the 
cross section of the accretion column at the neutron star surface has the shape of a thin annular arc, located at an off-center position with respect to the magnetic pole. BS76 quote a ratio of width to length of the arc of 0.025.
 The off-center position of the accretion spot has been confirmed qualitatively by magneto-hydrodynamic simulations of Romanova et al. (2004), however, the width of the arc is much larger compared with the idealized model of BS76, namely $\sim0.3-0.5$.  The simulations of Romanova et al. (2004) have been done for relatively small  magnetospheres, like those of accreting millisecond pulsars. Unfortunately, they could not be performed for the large magnetospheres of slowly spinning neutron stars with high magnetic fields due to limitations of computing power (Romanova, private communication). We assume that the picture of Romanova et al. (2004) is valid qualitatively also in our case. In  Fig. 4b we sketch the footprint of the column at the neutron star surface.
 This geometry leads to a modulation of the radiating area (and the X-ray flux) with the rotational phase at half the neutron star period. The ratio of minimum to maximum flux is given by $d_0/l_0$.
 Due to the opening of the magnetic field lines, the length and the width 
of the accretion column
scale with $r$ as $l (r) = l_{0} \times (r/R)^{3/2}$ and $d (r) = d_{0} \times (r/R)^{3/2}$.

The effective Eddington luminosity in the accretion funnel, near the
neutron star surface, is Becker et al. 2012

\begin{eqnarray}
L_{E}^{eff} & = & L_{E} \frac{l_0 d_0}{4\pi R^{2}}
\frac{\sigma_T}{\sigma_{||}} = \frac{GMm_{p}c}{\sigma_T}
 \frac{l_0  d_0}{R^2}   \frac{\sigma_T}{\sigma_{||}}
\nonumber
\\
\nonumber
 & = &  5.4\times10^{32}  \left(\frac{l_0}{10^{4}cm}\right)
\left(\frac{d_0}{6\times10^{3}cm}\right) \left(\frac{M}{1.4M_{\odot}}\right) \times
\\ 
 & &  \left(\frac{12.5 km}{R}\right)^{2} \frac{\sigma_T}{\sigma_{||}}\; \rm{erg\,s^{-1}}
%\eqnum{(1)}
\end{eqnarray}

where $L_E$ is the Eddington luminosity, $M$  is the mass of the
neutron star, $\sigma_T$ is the Thomson cross
section, and $\sigma_{||}$ is the average over energy cross section for photons travelling along the magnetic
field.

Since the flow in the accretion funnel is supercritical, a radiative
shock is established at a height $H$
above the neutron star surface (BS76, LS82, BW07). The upscattered photons in the shock escape
sideways, as a fan beam, and their propagation is controlled by Thomson scattering. The
transverse optical depth to electron scattering at or above the shock along the width of the funnel
is

\begin{eqnarray}
\tau_t = n_{e}(r) \sigma_T d(r) &  = & 
{\frac{\dot M}{m_p v_{ff}(r) l(r) d(r)} } \sigma_T d(r) 
\nonumber
\\
\nonumber
 & = & {\frac{\dot M}{m_p v_{ff}(r) l(r)    } } \sigma_T  
\\
 & = & {\frac{\dot M \sigma_T}{m_p v_0 l_0 }} {\frac{R}{r}}, 
\end{eqnarray}

where $n_{e}(r)$ is the electron density,  and $v_{ff}(r) =
v_0(R/r)^{1/2}$  the free-fall speed, 
 $v_0= (2GM/R)^{1/2} \approx 0.6c(M/M_0)^{1/2}(12.5 ~ {\rm km}/R)^{1/2}$, and $\dot M$ is the mass accretion
rate. The
transverse optical depth cannot be large, as it is in X-ray pulsars, because the BMC photons would
be downscattered before they escaped. On the other hand, it cannot be too small, because BMC
would be inefficient. In order to have a transverse optical depth of order a few, for accretion rates
of the order of $10^{15}$ g\,s$^{-1}$, the length $l_0$ must be of the
  order of $10^{4}$  cm.

The analytic solution for the spectrum produced at the shock, that was found by LS82, provides for
a high-energy power-law spectrum with index $\Gamma_{h} \sim 0.9$. Using a Monte Carlo code (Kylafis \&
Truemper, in preparation), we have obtained a similar spectrum, if the transverse Thomson optical
depth at the shock is $\tau_{t}(R+H)\sim5$ . Using eq. (2), this
implies that for $H<R$ the height of the
shock is given approximately by

$$
H \sim \left(\frac{\dot M \sigma_T}{5 m_p v_0 l_0} -1\right)R \sim 0.2R,
\eqno(3)
$$

where we have used $\dot M = 3\times10^{15}$ g\,s$^{-1}$ and $l_0 =
10^{4}$ cm. It is clear from the above, that despite the
low luminosity of AXPs and SGRs, the radiative shock may be located at some height, depending
on the accretion rate, the footprint of the accretion column, and the gravitational field.

For completeness, we mention that in the polar direction (i.e., parallel to the magnetic field), the
Thomson optical depth of the accretion column is
$$
\tau_p = \frac{2}{3} \tau_t(R) \frac{R}{d_0} >> \tau_{t}(R), 
\eqno(4)
$$

\subsection{Modeling the pulse shapes of 4U 0142+61}

As already noted above, after almost forty years of theoretical efforts, there is no fully selfconsistent
description yet of the radiation from an accretion column. This is true in particular for
the angle- and energy-dependence of the emission. The most advanced approach is the one by
BW07, which describes the spectral formation in terms of TC and BMC, but uses angle averaged
scattering cross sections, which are not suited to derive angular distributions. We therefore use a
pragmatic approach and introduce beaming functions to describe the angular dependence of the
different beams. This method has been used widely for the analysis of X-ray pulsar profiles (e.g.,
Leahy 1990, 2003; Kraus 2001).

To model the observed pulse profiles, we assume that the neutron star has a centered magnetic
dipole. Thus, the fan beam, which escapes sideways from the column, and the polar beam are
orthogonal. The beaming functions are given by  $f \sin^m \theta$  and $p \cos^n \theta$, for the fan beam and the
polar beam, respectively. Here $\theta$ is the polar angle, i.e. the angle between the direction of
photon propagation and the magnetic dipole axis. The amplitudes $f$
and $p$ and the exponents $m$
and $n$ are free parameters, which may be different for different
energy intervals. Note that $m=1$
or $n=1$ corresponds to the case of isotropic emission, e.g. blackbody emission. The
ansatz for the beaming function is a generalized version of the angular distributions of photons
presented by Nagel (1981) and Meszaros \& Nagel (1985a,b) for columns and slabs, which can
be described by $\sin^m \theta$ and $\cos^n \theta$, respectively, with  $m$ and $n \leq 1.5$, except for an energy
interval around the cyclotron resonance. In that work, the full angle- and energy-dependent
magnetic scattering cross sections, including effects of vacuum polarization were used, assuming a
homogeneous temperature and density distribution in the radiating slab or column. However, since
in the outer layers of a slab or a column the plasma temperature will drop towards the surface,
photons escaping perpendicular to the surface will escape from deeper layers than photons escaping
in other directions. This leads to a limb-darkening effect, which makes the beaming patterns
narrower compared with those of Nagel (1981) and Meszaros \& Nagel (1985a,b). Thus, the
expected exponents of the beaming will be significantly larger than 1.5.

A model with orthogonal fan and polar beams will lead to symmetric pulse profiles with double
pulses separated by 0.5 in phase, in disagreement with the observed pulse profiles, which show
asymmetries and slightly different pulse separations. To account for these effects we assume, as
discussed in Section 3.2, that the accretion column has a non-cylindrical  cross
section  with major axis $l_0$  and minor axis $d_0$ (c.f. Fig 4b).

To account for gravitational bending, we use the expression derived by
Beloborodov (2002). For a photon emitted near the neutron star at an
angle $\theta$ with respect to the
radial direction, the angle $\theta + \delta$, at which it arrives at infinity, is given by

$$
 1 - \cos(\theta + \delta) = \frac{1 - \cos\theta}{1-\frac{R_s}{R}},
\eqno{5}
$$

where $R$ is the radius of the neutron star and $R_s = 2GM/Rc^2$, is its Schwarzschild radius. This
equation provides a very good approximation in a Schwarzschild metric
for $R/R_s > 2$. We
assume $M = 1.4 M_\odot$ and $R=12.5$ km, viz. $R/R_s = 3$ . An
important parameter is the height $H$ of
the radiative shock above the neutron star surface. In the following
we discuss three cases: $H=0$, $H=2$\,km, and  $H=0.5 R$.

a)  $H = 0$: This choice is motivated by the general view that at low accretion rate the matter is
stopped by the radiative shock at a position close to the stellar surface. The fitting parameters are:
the inclination $i$ of the system, the angle $\alpha$ between the rotation axis and the magnetic dipole
axis, the axis ratio $d_0 / l_0$ , the azimuthal angle $\phi$ of the long axis, phase zero of the pulse
profile, the exponents $m$ and $n$ of the beaming functions for the fan and the polar beam, and the
normalizations $f$ and $p$ of the fan beam component and the polar beam component,
respectively. We remark that the first five parameters ($i,\; \alpha, \;
d_0/l_0, \; \phi,$ and phase zero) are fitted
jointly for all energy bands, while the others are fitted for each energy band separately.

The fit is done by a hierarchical grid search. In an outer loop the global parameters are
scanned in a five-dimensional grid. In an inner loop the two exponents and the flux ratio between
the fan and polar beam are scanned in a three-dimensional grid, for each combination of the
global parameters of the outer grid. For each candidate pulse profile, the overall normalization of
the flux is directly determined by the ratio of the mean flux of that profile and the measured
profile in the corresponding energy range.

This procedure is applied several times, starting with a coarse grid covering a wide range of
parameters. This grid is subsequently refined by a denser grid around the solution found in the
preceding step. We use the $\chi^2$ statistics as the criterion for the agreement between the
observed and calculated pulse profiles. In order to avoid that the solution is dominated by the
low energy profiles, which have the best statistical quality, we add systematic errors to the
statistical errors, individually for each energy band, so that formally acceptable fits are obtained.
This method allows us also to quantify to which accuracy our model is capable of reproducing the
observed pulse profiles. We find systematic errors of 2.1\%, 1.7\%, 4.4\%,
3.5\%, 3.5\% for the pulse profiles of the corresponding ranges 0.8 -- 2.0 keV, 2 -- 4 keV, 4 -- 8 keV,
8 -- 16.3 keV, 20 -- 50 keV (for the 50 -- 160 keV band, no additional error is required). We then estimate the uncertainty in the model parameters 
by subsequently varying the individual
parameters in both directions (while keeping the other parameters fixed at the best-fit values), until the $\chi^2$ increases by 1, 
which corresponds to the 68\% confidence
level for 1 interesting parameter. 
Fig. 5a shows the model fit to the observed light curves. The
inclination is $i=42.1 \pm 0.4$
degrees, the angle between the dipole with respect to the spin axis
$\alpha=19 \pm 0.9 $ degrees and the axis
ratio of the accretion column is $d_0/l_0=0.69 \pm 0.01$, where the errors are
$1 \sigma$.

Our model gives us information about the full beaming geometry, and thus allows us to determine
the ``intrinsic luminosities'' of the different beams. This is a major advantage compared to
analysis methods which are purely based on the observed flux, i.e., the flux which we sample
along the line of sight during the rotation of the neutron star. Depending on the geometrical
circumstances, the {\textit{observed flux may not be representative for the X-ray emission of the neutron
star}}. In the following we utilize the information about the full beaming geometry to compute
which fraction of the intrinsic fan beam luminosity irradiates the neutron star and compare this
fraction with the intrinsic luminosity of the polar beam. For clarity, we focus in the following on
the properties of one pole (due to the symmetry of our model, the properties of the other pole are
identical).

According to the $H=0$ model, the intrinsic 0.8 -- 160 keV luminosity of the fan beam
is $4.4\times10^{35}$ \ergs,
which includes the fraction which falls onto the neutron star and the one which escapes. The
intrinsic luminosity of the polar beam is $0.7\times10^{35}$\ergs, viz. ~ 15 \% of the energy of the
intrinsic fan beam. The expected fraction of the total luminosity of the fan beam hitting the
neutron star is 50\%. Thus the neutron star reemits only a fraction 15/50= 30\% of the incident flux,
while the rest disappears, e.g. in the neutron-star crust. As this seems quite unlikely, we
investigated an alternative scenario, where the radiative shock is not located at the stellar surface,
but at some height $H$ above it. In this case, the neutron star intercepts less than half of the incident fan beam
flux.

b) $H=2$\,km: In order to compute which fraction of the total fan beam luminosity reaches the
neutron star for the case $H>0$, it is necessary to know the trajectories of the emitted photons in
the presence of gravitational bending. Such trajectories are found in the work of Leahy 
(2003) for a neutron star radius $R=12.5$\,km. He shows that X-rays emitted from the accretion
column at a height $H$ reach the neutron star if they are emitted at
zenith angles $\theta > \theta_i(H)$, while
the rest escapes. Implementing this condition into the model and
assuming $H=2$\,km, the solution
is shown in Fig. 5b. For this model, we find systematic errors of 2.0\%, 3.1\%, 7.1\%, 6.1\%, 7.5\% 
for the pulse profiles of the corresponding ranges 0.8 -- 2.0 keV, 2 -- 4 keV, 4 -- 8 keV, 8 -- 16.3
keV, 20 -- 50 keV (for the 50 -- 160 keV band, no additional error is
required). The inclination is $i=59.6\pm2.4$
degrees, the angle between the dipole and  the spin axis
$\alpha=31.2\pm1.7$ degrees, and the
axis ratio of the accretion column is $d_0/l_0 = 4/5$. The intrinsic
 0.8 -- 160\,keV luminosity of the fan beam is $1.8\times10^{35}$\ergs\ and that of the
polar beam is $0.3\times10^{35}$\ergs. Thus the ratio of the luminosities
is again ~15\%. However, the fraction of the total luminosity of the fan beam which reaches the
neutron star is now smaller. Its exact value depends on the exponent
$m$. For a typical $m=4$, it is
about 20\%. In this case, an X-ray albedo of 15/20 = 75\% would be required for the neutron star,
which looks more plausible than the 30\% found for $H=0$.

c) $H = 0.5 R$: We also investigated the case H = 6.25\,km and obtained fit
 results ($i = 55.0$\,deg, $\alpha = 30.5$\,deg,  $d_{0}/l_{0} = 4/5$), which are very similar
 to those for $H = 2$\,km. This is a consequence of the fact that the effect of
 gravitational bending becomes less important with increasing height.
 Compared to $H = 2$\,km, the intrinsic luminosity of the fan beam increases
 slightly (by less than 3\%), while the polar beam luminosity stays the same, so that the ratio is again $\sim15\%$. At $H = 0.5 R$, however, the neutron star intercepts only $\sim5\%$ of the incident fan beam flux, so that the fan beam could power only $5/15 = 1/3$ of the polar beam luminosity.
  
 The fact that the fit results are rather insensitive to $H$, for $H$ between 2 km and 6.25 km, allows us to compute the maximum value of $H$ in a straightforward way, by determining the height at which the neutron star would intercept 15\% of the fan beam luminosity. This value depends on the exponent $m$ and decreases with increasing $m$. For representative values $m = 2$ and 3, it is $\sim3.5$\,km and $\sim2.8$\,km, respectively. We note that the upper limit of $\sim3.5$\,km or 0.3 R, which we find for the height of the radiative shock, is consistent with the height estimate $(\sim2.5 ~ \rm{km})$ presented in section 3.2.

\section{The phase-dependent spectra - an estimate of the magnetic 
dipole field of AXP 4U 0142+61?} 

As already discussed in Sections 3.2 and 3.4 due to the beaming geometry and enhancement by
gravitational bending, a substantial part of the fan beam hits the polar cap photosphere and will
heat it up. A fraction of the infalling photons will be scattered, leading to a reflected beam emitted
in the polar direction. The reflection coefficient depends on photon energy and reaches a
maximum at the local cyclotron frequency. Thus, one expects that the intensity ratio of the polar to
the fan beam increases towards the cyclotron energy. Such an increase is indeed observed (Fig. 3).
At $\sim60$ keV, the polar beam reaches its maximum, followed by a decline or cutoff beyond $\sim70$
keV. Thus, the corresponding cyclotron energy is $\sim60$ keV, implying  a polar magnetic field of
$B \sim 5 \times 10^{12} (1+z)$  G, where $z$ is the gravitational redshift. We note that the rather large apparent
width of this feature may be due to the variation of the magnetic dipole field over the polar cap
photosphere, which according to observations has a radius of $\sim8$ km (White et al. 1996; Israel et
al. 1999; Juett et al. 2002). Clearly, a confirmation of this spectral feature and a more accurate
determination of its position and width in energy will allow us to further test the accuracy of these
estimates.

\section{Heating of the polar cap}

It is well known that the polar caps of AXPs/SGRs show significantly higher photospheric
temperatures compared with other relatively young neutron stars, like radio pulsars (e.g. Aguilera
et al. 2008). It is evident that our model provides a natural explanation for this observational fact,
since it predicts a strong illumination of the photosphere surrounding the accretion column by the
fan beam. The efficiency of this process is increased by gravitational
bending.

Because of the relatively high threshold ($>0.8$\,keV), the XMM-Newton data of den Hartog et al.
(2008) taken in 2003 and 2004 do not contain information on the blackbody component. The
Chandra MEG and HEG observations of 4U 0142+61 on 23 May 2001 (Juett et al. 2002) have
measured the spectrum at lower energies (0.5 - 10 keV) which can be fitted by a blackbody
($kT=0.418\pm0.013$ keV, $L_{bb} = 5.7\times10^{34}(d/3.6\rm{kpc})^{2}$ \ergs) and a power-law component with a
photon index $\Gamma_s = 3.3$. The observed energy flux in the 0.8 --  2 keV interval is 47.5 $\rm{eV\,cm^{-2} s^{-1}}$.
In our data, based on the XMM-Newton observations, the corresponding flux is 87.6 $\rm{eV\,cm^{-2} s^{-1}}$.
The difference may be due to source variability and/or systematic errors. Assuming that the ratio of
blackbody and power law energy flux (which is $~ 1$) is the same in both data sets we find a
blackbody luminosity $L_{bb} = 10^{35}$\ergs\ expected for the XMM-Newton spectrum.

The blackbody luminosity of 4U 0142+61 can also be derived from data presented by Enoto et al.
(2011). They fit two blackbodies to the soft component with $kT_1 =
0.337$ keV, $R_1 = 13.2$ km and $kT_2=0.633$ keV, $R_2 = 2.3$ km. The
corresponding blackbody luminosities are   $L_{1} =
7.25\times10^{34}$\ergs\ and  $L_{2} = 2.69\times10^{34}$\ergs. Thus, the total blackbody luminosity is  $L_{bb} = 10^{35}$\ergs, in good
agreement with the value derived from the Chandra data for the time of the XMM-Newton
observations. This result, together with the agreement between the XMM-Newton data and the
Suzaku data presented in table 1, shows that the observational data are consistent.

Traditionally, the soft component of AXPs and SGRs has been fitted by two different models: a)
Blackbody plus power law (BB+PL) or b) double blackbody (dBB). Halpern \& Gotthelf (2004)
discussed the merits of both models and supported interpretation b. Our results shed new light on
this question. In our model, the power-law component can be attributed to thermal Comptonization,
taking place in the hot plasma below the radiative shock. However, a natural consequence of our
model is that the temperature falls off on the polar cap with the distance from the accretion column,
which would be consistent with the double blackbody fit discussed by Enoto et al. (2011). It seems
that at present the observational data and the model predictions are not good enough to distinguish
between these alternatives or their combination.

\section{Comparison with X-ray binary neutron stars}

It is an interesting question why these AXPs/SGRs have hard X-ray tails up to 100 keV or more
(Enoto et al. 2010), while in general the accreting binary neutron stars show an exponential cutoff
at energies of 10 -- 20 keV. The major difference between the two classes is that AXPs/SGRs have
luminosities which are about two or three orders of magnitude lower than typical X-ray binaries
like Her X-1. To our knowledge, the only known accreting X-ray pulsars in binary system, which
come close to 4U 0142+61 with respect to luminosity and magnetic field strength, are 4U
0352+309 (X Persei) and 4U 2206+ 54, which we discuss below. 
Although there are a few additional 
sources that could be considered as members of this class of objects (e.g. 4U 1145-61 and 4U 1258-61), the 
scarce available hard X-ray data on them do not allow us to investigate 
the nature (and significance in the case of 4U 1258-61) of their hard X-ray emission at the level of detail that is possible for
X Persei and 4U 2206+ 54.

4U 0352+309 (X-Per): According to BeppoSAX observations of di Salvo et al. (1998), this source
has a luminosity of  $2.4 \times 10^{34}$\ergs, with half of the luminosity in the 0.1 -- 10 keV band and the
other half in the 10 - 100 keV band. A crude estimate derived from the spin-up rate gives a
magnetic dipole field for 4U 0352+309 of $B\sim 2.5\times10^{12}(1+z)$\,G (Ghosh \& Lamb 1979). The
detection of a cyclotron line at $\sim29$ keV, implying a magnetic field
strength of $B\sim2.5\times10^{12}(1+z)$\,G, has been reported by Coburn et al. (2001). Thus, the magnetic field
strength of 4U 0352+309 (X Persei) is comparable with that expected for fallback-disk accretors.
The spectrum of the source exhibits a soft thermal component below 20 keV and a distinct hard
tail (di Salvo et al. 1998). The photon number spectral index of the hard component between 30
and 70 keV is $\Gamma_h \sim1.35$, while for the hard tail of 4U
0142+61 it is $\Gamma_h \sim0.93$  (den
Hartog et al. 2008, see also Truemper et al. 2010) or $\Gamma_h \sim0.9$  (Enoto et al. 2011). Thus, this
source has very similar spectral characteristics as 4U 0142+61.
 Recently, Lutovinov et al. (2012) described the activity of the source during 2001-2011 and
reported several consecutive outbursts during which the luminosities
increased up to $1.2 \times 10^{35}$\ergs\ and a spectrum showing a
distinct hard X-ray tail up to energies of $\sim160$ keV. The cyclotron line at 29 keV is confirmed.

We suggest that the power-law tail of 4U 0352+309 is due to the same mechanism operating in 4U
0142+61, namely bulk-motion Comptonization.\footnote{We note, that during the refereeing
process of our work, Doroshenko et al. (2012) published a paper in which they analyzed the
spectrum of 4U 0352+309 observed with Integral and fitted it with a semi-phenomenological
thermal and bulk-motion Comptonization model for the soft and hard
component, respectively.}

4U 2206+54: This is a pulsar showing very slow pulsations ($P=5560$\,s, Reig et al. 2009)
accreting matter from the stellar wind of an O-type companion. A detailed analysis of the magnetorotational
evolution requires a magnetic field of the neutron star $B>5\times10^{13}$\,G (Ikhsanov \&
Beskrovnaya 2011). This is in conflict with a possible cyclotron resonance scattering feature at $\sim30$\,keV claimed by Torrejon et al. (2004), Masetti et al. (2004) , Blay et al. (2005), and Wang
(2009), which would indicate a magnetic field strength of $\sim3.3\times10^{12}$\,G, but which could not be
detected by other observations (Reig et al. 2009; Wang 2010). At any rate, the dipole magnetic
field must be large, although it will be smaller than the magnetic fields usually assigned to
magnetars (Reig et al. 2012). The source shows intensity variations of a factor 5 in the soft band (2
-- 10 keV), ranging from 1.5 to $8\times10^{-3}$ $\rm{photons\, cm^{-2} s^{-1}}$ . The average luminosity in the 2 -- 10 keV
band is $1.5\times10^{35}$\ergs\ for a distance of 2.6 kpc (e.g. Reig et al. 2012). Thus the luminosity and
the magnetic dipole strength are comparable with those of AXPs/SGRs.

But there are also other similarities: 1) The soft spectrum can be modeled by a blackbody and a
power-law component. The blackbody radius is $\sim2.6$ km which is consistent with the size of a hot
polar cap of the neutron star (Reig et al. 2012). 2) At energies $> 20$ keV a power-law spectrum is
observed which has been measured up to energies of more than 100 keV whose steepness depends
on the intensity level. At low luminosity levels 
($\gtrsim 10^{34}$\ergs), the source shows a hard power-law
tail at energies > 20 keV ranging up to more than 100 keV with a slope
$\Gamma_h \sim 2$ (Torrejon et al.
2004; Masetti et al. 2004; Reig et al. 2009). At higher luminosity levels ($\sim2.7\times10^{35}$\ergs) the
hard spectrum becomes significantly steeper with $\Gamma_h \sim 3$  (Reig et al. 2012). Actually its overall
shape approaches that of high luminosity pulsars like Her X-1 which shows a spectral cut-off at 
$\sim20$ keV.

Concluding this discussion, we note that the similarity in the spectral behavior of 4U 0352+309 and
4U 2206+54 compared with AXPs/SGRs on the one hand and the difference from the majority of
the high luminosity binary X-ray sources on the other, supports our view (see section 3.2) that in
the luminosity range around $10^{35}$\ergs\ the transverse optical depth of the accretion column
($\tau\sim5$ ) is favorable for the production and escape of BMC photons.

\section{Summary and conclusions}

\begin{enumerate}

\item
We have shown in a semi-quantitative way that bulk-motion/thermal Comptonization in an
accretion column, formed by a dipole magnetic field of strength $B\sim 10^{13}$ G, describes well not
only the soft and hard X-ray spectra but also the phase-dependent energy spectra of AXP 4U
0142+61.

\item
Furthermore, we have shown in a quantitative way that the model explains naturally the
observed pulse profiles in all X-ray bands. The energy-dependent pulse profiles and their
constancy over long periods of time constitute significant observational constraints for the
proposed models. Our model explains them in a simple and natural way by the formation of two
``beams'', one perpendicular to the accretion column (main pulse) and another parallel to it
(secondary pulse).

\item
Our beaming model takes the gravitational bending and a noncircular shape of the accretion
column into account, allowing us to determine both the angle $a$
 between the spin axis and the 
 magnetic dipole axis, and the inclination $i$ of the spin axis with respect to the line of sight.
Assuming that the radiative shock is located very close to the stellar surface ($H=0$), we find a
good fit with $i\sim40$ degrees and $\alpha \sim 20$\,degrees. However, the ratio of polar to fan beam flux is
too small, as only 30\% of the impinging fan beam flux is converted into the scattered and thermal
albedo. For a height of the radiative shock of $H=2$ km, we find $i
\sim 60$ degrees, $\alpha \sim 30$ degrees,
and a more reasonable value for the albedo, which is $\sim75$\%. We conclude that our fits provide
evidence for a height $H$ of the order of at most 3.5 km.

\item
The Suzaku observations (Enoto et al. 2010) have provided a homogeneous set of spectra of 2
SGRs and 5 AXPs, showing that the fluxes of the soft $F_s$ and the
hard $F_h$ spectral components are
roughly equal, with variations of a factor
3. In view of the similarity of their broad-band spectra, it
is no surprise that accretion models with BB, TC, and BMC components can fit their broad-band
(0.5 - 200 keV) spectra (Zezas et al., in preparation). The observed ratio of these observed
fluxes $F_h / F_s$ will depend on the range of polar angles $\theta$ swept by the line of sight during one
revolution. In addition, one expects in our model that the hard luminosity produced by BMC
correlates with the accretion rate, and for disk accretion the latter is expected to correlate with the
period derivative $\dot P$. Thus, one expects a correlation of $F_h /
F_s$ with $\dot P$. This is in agreement with
the Suzaku data which show a correlation of  $F_h / F_s$  with $\dot P$.

\item
We suggest that the formation of hard X-ray spectral tails in accretion columns depends
crucially on the optical depth of the accretion column, which is expected to correlate with X-ray
luminosity. For high luminosity sources, the transverse optical depth
$\tau_t$ of the accretion column is
high, and the high energy photons produced by BMC are thermalized before they escape. In
AXPs and SGRs, having luminosities a factor of $\sim100$ less, the optical depth across the column
is large enough to ensure efficient upscattering of photons to high energies ($E \gtrsim 100$ keV) and low
enough to allow their escape before thermalization. We suggest that
this requires $\tau_t \sim 5$.

\item
The fact that the accreting X-ray binary 4U 0352+309 (X Per), having a luminosity about a
factor of 10 smaller than 4U 0142+61, shows a luminous hard tail strongly supports our model. In this case, the optical depth may be expected to be even smaller, which
would explain the relatively low cutoff of the hard component ($\sim 65$ keV). The low luminosity
accreting neutron star 4U 2203+54 has a hard tail as well showing a significant steepening of its
hard spectrum at transitions from low to high luminosities, which is in line with the conclusion of
point 5.

\item
It is well known that the polar caps of AXPs/SGRs show significantly higher photospheric
temperatures compared with other relatively young neutron stars, like radio pulsars or
X-ray dim neutron stars (XDINs) (e.g. Aguilera et al. 2008). Evidently, our model provides a
natural explanation for this observational fact, since the strong illumination of the photosphere by
the fan beam acts as an additional heat source. The efficiency of this process is increased by
gravitational bending. The heat influx into the photosphere will depend primarily on the power of
the fan beam and its width, as well as on the gravity of the neutron
star.

\item
We propose that the spectral bump of the secondary pulse (which corresponds to the polar
beam) is caused by cyclotron resonance reflection of the fan beam by the magnetized photosphere.
This leads to magnetic field strength of $B\sim 5\times10^{12}(1+z)$ G, which falls in the range of dipole
magnetic fields ($10^{12}-10^{13}$ G) attributed to AXPs/SGRs in the framework of the fallback-disk
scenario. Support for the above proposal is the steepness of the polar beam spectrum in the 10 -- 60 keV range.  

\item
We stress that the few giant bursts observed from SGRs cannot be explained by any accretion
process. The same is true for other bursts with large super-Eddington luminosities. They are most
likely caused by processes taking place at the stellar surface, e.g. conversion of super-strong
magnetic fields ($B \sim 10^{15}$ G) into radiation triggered by crustal shifts, as discussed in the classical
magnetar literature (e.g. Thomson \& Duncan 1995). Our analysis suggests that these events do not
take place in the dipole field, but in localized multipole fields. This situation is qualitatively
similar to that of the Sun, which shows flare activities in sunspot fields, which are larger than the
solar dipole field by at least two orders of magnitude.

\item
Finally, we note that the kind of analysis presented in this paper - applied to other
AXPs/SGRs observed with sufficiently large photon statistics - can be used to study in detail the
radiative properties of accretion columns and to narrow the gap between observations and
theoretical models in this important field of Astrophysics.

\end{enumerate}

\acknowledgements
This research has been supported in part by EU Marie Curie project no. 39965, 
EU REGPOT project number 206469 and by EU FPG Marie Curie Transfer of Knowledge 
Project ASTRONS, MKTD-CT-2006-042722. \"{U}.E. acknowledges research support 
from T\"{U}B{\.I}TAK (The Scientific and Technical Research Council of Turkey) 
through grant 110T243 and support from the Sabanc\i\ University Astrophysics 
and Space Forum.   We also
thank Xiao-Ling Zhang (MPE) for her help with a search of the ISGRI data bank.

{\it Facilities:} \facility{INTEGRAL},  \facility{XMM-Newton}.

\newpage

\begin{figure}
%\centering
\rotatebox{270}{\includegraphics[angle=0,width=12cm]{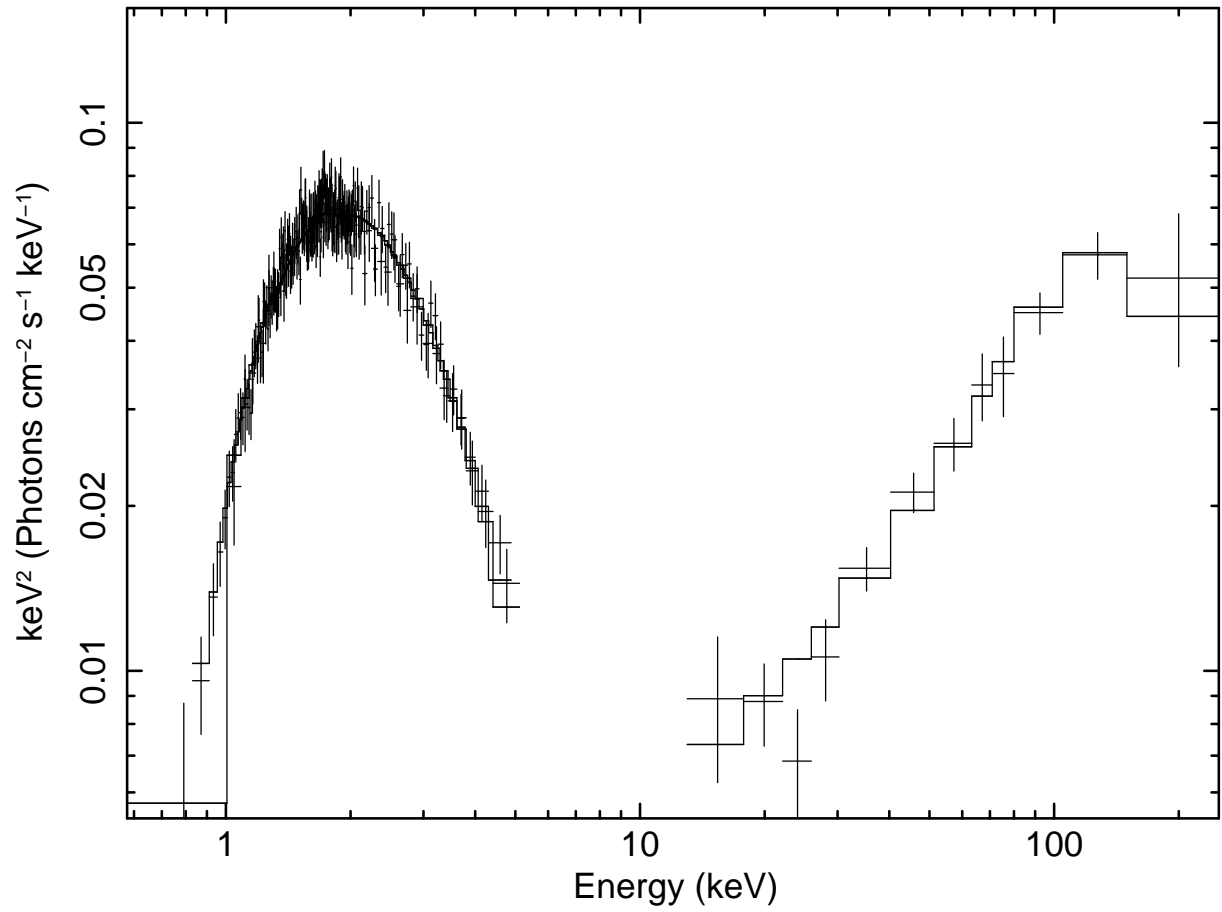}}
\caption{ \label{Fig1}
The best-fitting COMPTB model, along with the pulse phase averaged  data from Chandra MEG and HEG (< 10keV) and Integral ISGRI (> 15 keV), c.f. Truemper et al. (2010.} 
\end{figure}

\begin{figure}
%\centering
\includegraphics[angle=0,width=12cm]{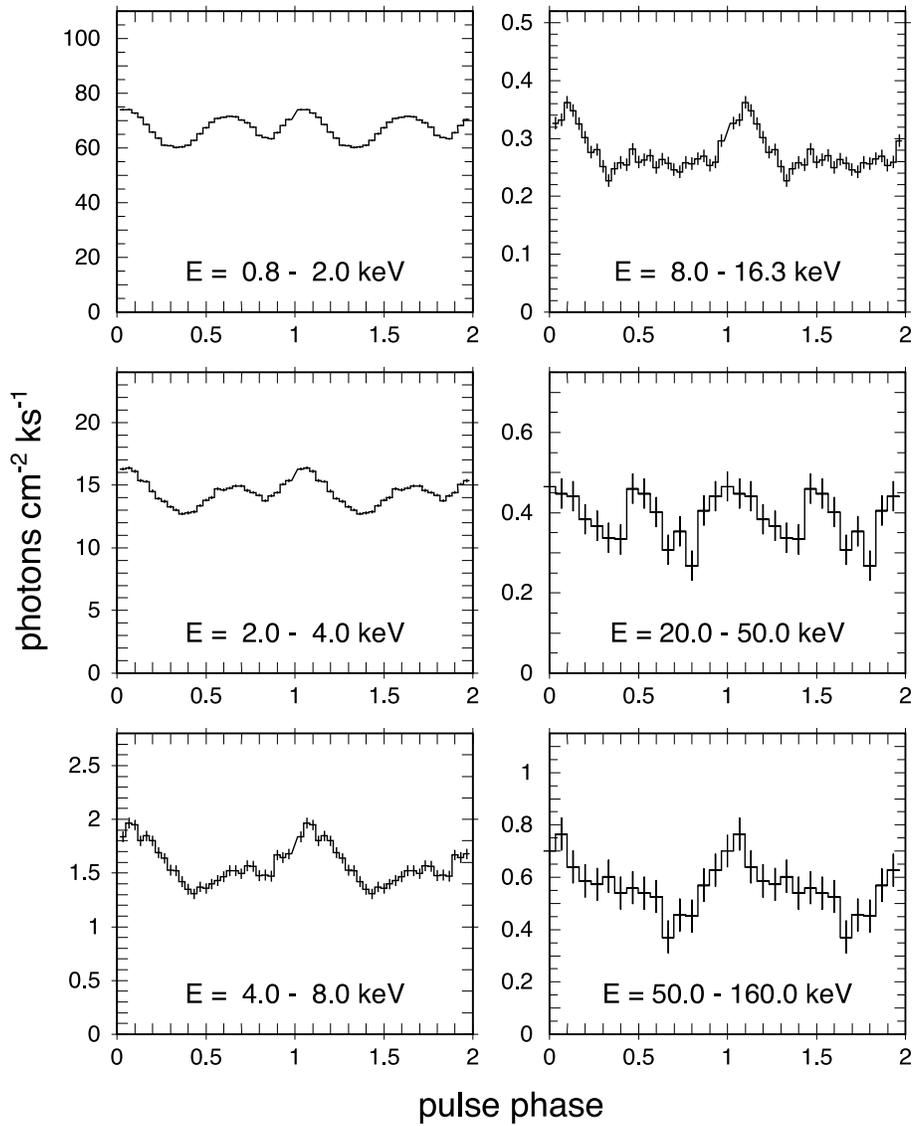}
\caption{ \label{Fig2}
The histograms show the total pulse profiles 
(pulsed plus non-pulsed components)
of 4U 0142+61, derived from the 
observational
data of den Hartog et al. (2008). 
}
\end{figure}

\begin{figure}
%\centering
\includegraphics[angle=0,width=12cm]{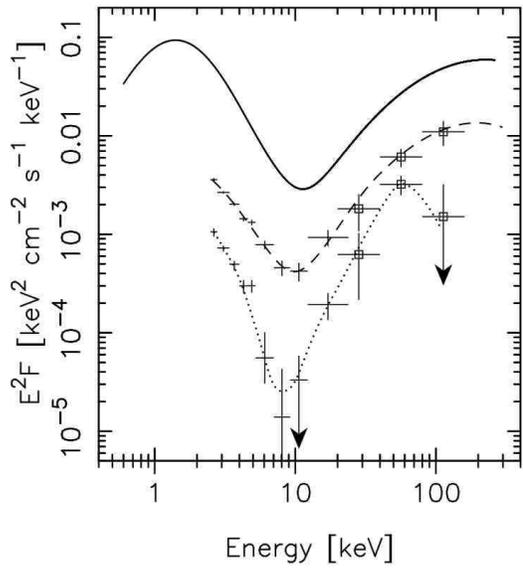}
\caption{ \label{Fig3} 
This Figure is taken from Fig. 8 of den Hartog et al. (2008).  The black line (top one) represents the INTEGRAL/XMM-Newton total spectrum fit shown in that paper.  The spectrum of the main pulse (in our model the fan beam), taken from the phase interval 0.85 -- 0.35, is shown as a dashed line).  The spectrum of the secondary pulse (in our model the polar beam), taken from the phase interval 0.35 -- 0.85, is shown as a dotted line.  The spectrum of the polar beam shows a steep rise up to a maximum at $\sim 60$ keV.  In our model, this bump is tentatively interpreted in terms of enhanced reflection due to cyclotron scattering of fan beam photons at the polar photosphere.
}
\end{figure}

\begin{figure}
%\centering
\includegraphics[angle=0,width=15cm]{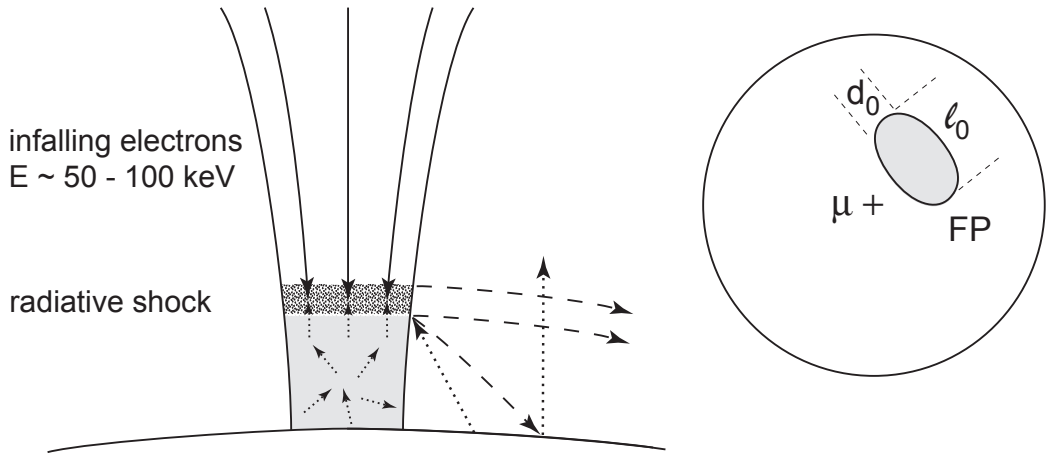}
\caption{ \label{Fig4} 
Fig. 4a (left): Schematic description of the beaming pattern. The different 
arrows show the mean direction of the respective component. Dashed
arrows: Fan beam consisting of photons produced by thermal
Comptonization ($E< 10$ keV) and bulk-motion Comptonization ($E>10$  keV).  It
is subject to gravitational bending and part of the fan beam hits the
neutron star photosphere, where it is scattered or absorbed.  Dotted
arrows: Polar beam consisting of fan beam photons after reflection
(scattering) from the photosphere. 
Also thermal photons emitted from the hot photosphere contribute to the polar beam.
\newline
Fig. 4b (right): Footprint (FP) of the accretion column at the neutron star surface (see text).
}
\end{figure}

\begin{figure}
%\centering
\includegraphics[angle=0,height=14.5cm]{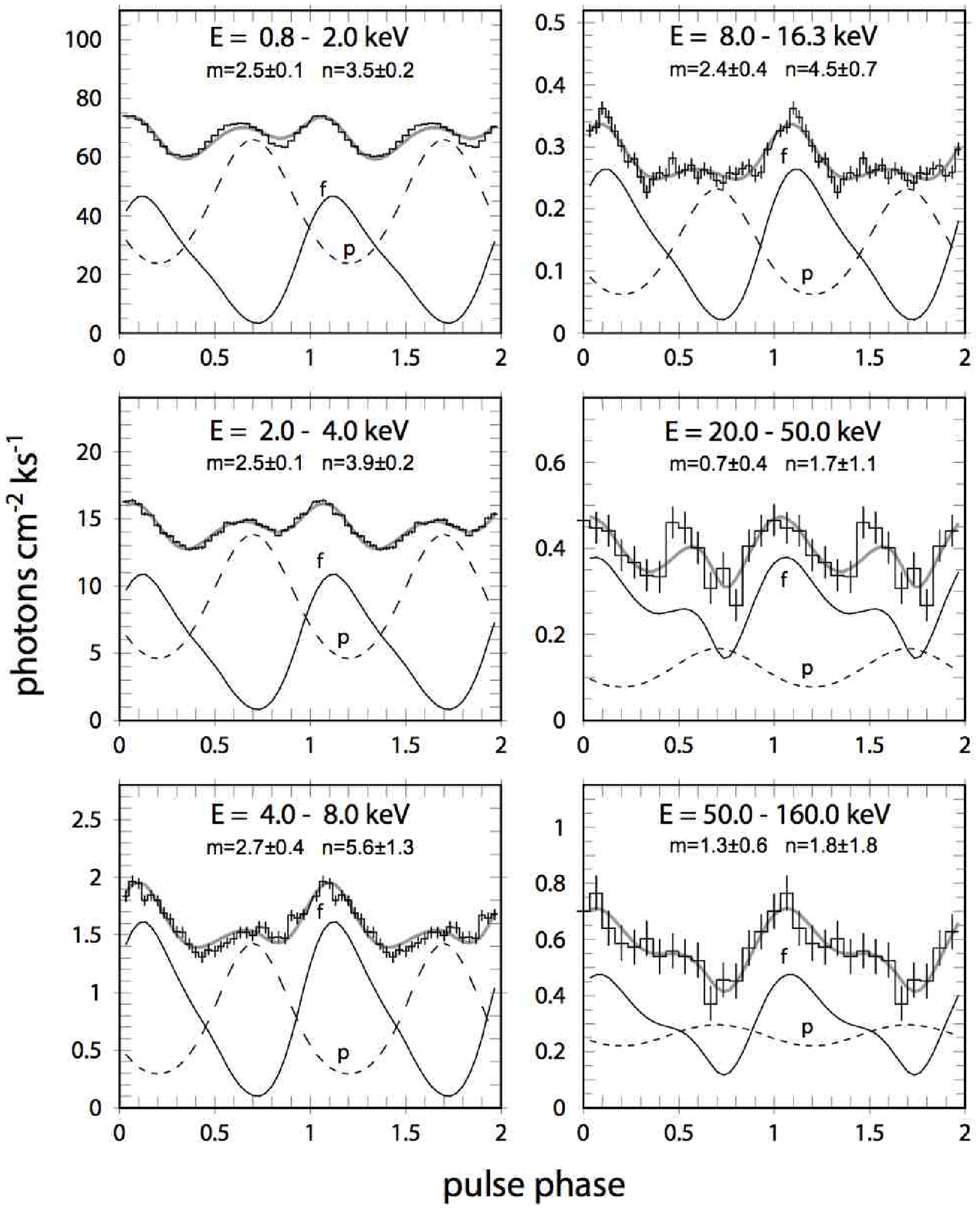}
\includegraphics[angle=0,height=14.5cm]{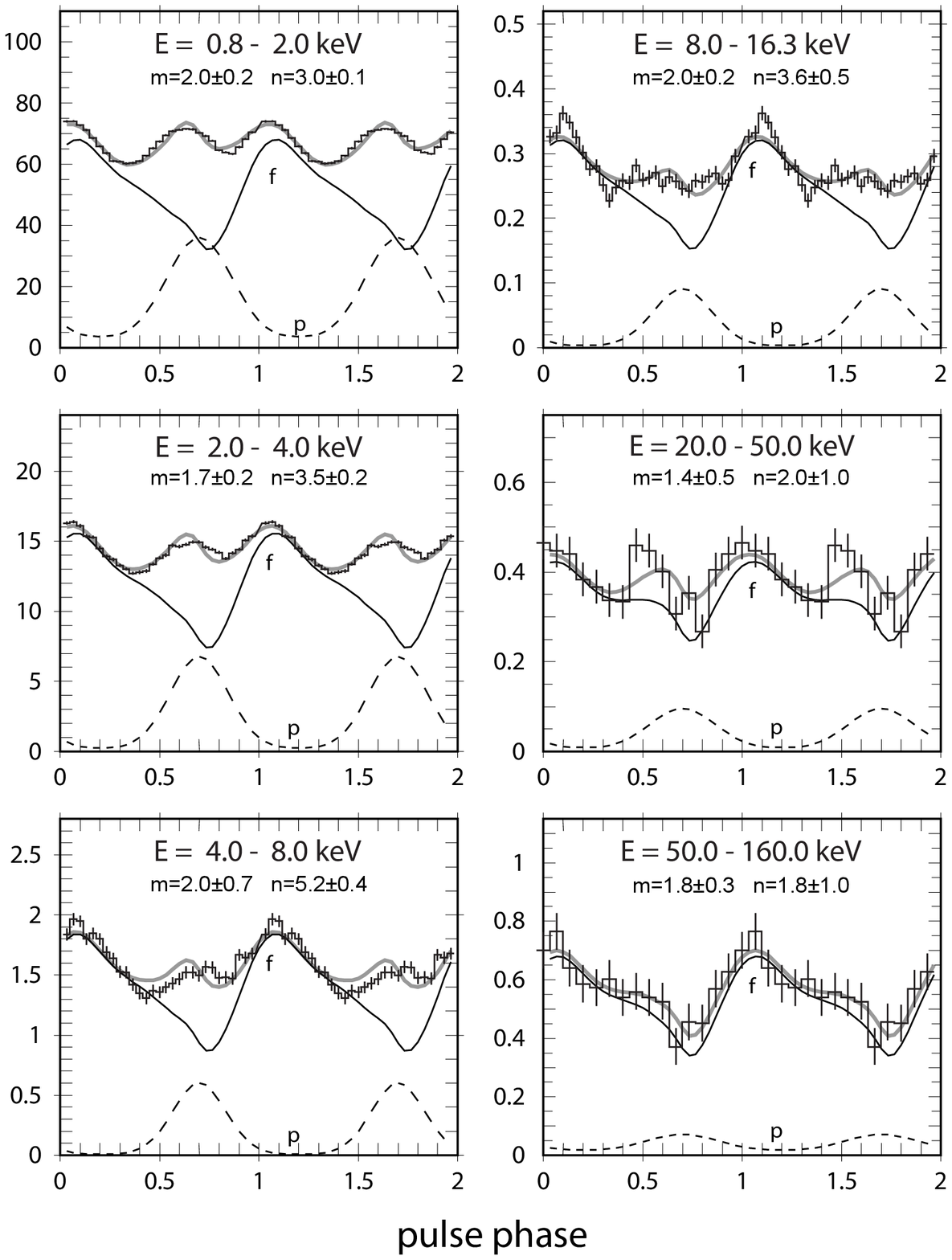}
\caption{
The black histograms show the total observed pulse profiles (pulsed plus
non-pulsed components) of 4U 0142+61 as given in Fig. 2.  The best-fit model for the
pulse profiles is shown as a thick grey curve, in 5a for a height of the radiative
shock at  $H=0$, and  in 5b for  $H=2$\,km. The decomposition of the pulse profiles into the fan  beam  and the polar beam components are shown as solid and dashed lines, respectively. $m$ and $n$ are the exponents of the beaming functions of the fan beam and the polar beam, respectively (see text). 
}
\label{Fig5}
\end{figure}

\end{document}